# On the Use of Graphene to Improve the Performance of Concentrator III-V Multijunction Solar Cells

Laura Barrutia[1], Iván Lombardero[1]*, Mario Ochoa[1], Mercedes Gabás[1,2], Iván García[1], Tomás Palacios[3], Andrew Johnson[4], Ignacio Rey-Stolle[1] and Carlos Algora[1]

[1]Instituto de Energía Solar – Universidad Politécnica de Madrid, ETSI de Telecomunicación, Avda. Complutense 30, 28040, Madrid, Spain

[2]The Nanotech Unit, Departamento de Física Aplicada I – Universidad de Málaga, 29071 Málaga, Spain

[3]Dept. of Electrical Engineering and Computer Science, Massachusetts Institute of Technology, Cambridge, MA (USA)

[4]IQE plc, Pascal Close, St. Mellons, Cardiff. CF3 0LW, United Kingdom

* *Corresponding author.* ivan.lombardero@ies.upm.es (I. Lombardero)

*Abstract.* Graphene has been intensively studied in photovoltaics focusing on emerging solar cells based on thin films, dye-sensitized, quantum dots, nanowires, etc. However, the typical efficiency of these solar cells incorporating graphene are below 16%. Therefore, the photovoltaic potential of graphene has not been already shown. In this work the use of graphene for concentration applications on III-V multijunction solar cells, which indeed are the solar cells with the highest efficiency, is demonstrated. Firstly, a wide optoelectronic characterization of graphene layers is carried out. Then, the graphene layer is incorporated onto triple-junction solar cells, which decreases their series resistance by 35% (relative), leading to an increase in Fill Factor of 4% (absolute) at concentrations of 1,000 suns. Simultaneously, the optical absorption of graphene produces a relative short circuit current density decrease in the range of 0-1.8%. As a result, an absolute efficiency improvement close to 1% at concentrations of 1,000 suns was achieved with respect to triple junction solar cells without graphene. The impact of incorporating one and two graphene monolayers is also evaluated.

## Introduction

Aiming at the reduction of the cost of Concentrator Photovoltaic (CPV) systems, the use of ultra-high concentration levels ($\geq$ 1,000 suns) in III-V multijunction solar cells (MJSCs) has been proposed [1-3] Although very high efficiency MJSC designs are being developed for this purpose –with current champion efficiencies around 44-46%–,[4-6] their series resistance is responsible for important efficiency losses when operating under highly concentrated light [7]. The most important components of the series resistance in a MJSC are those of the front metal grid and top cell emitter/window, given the fact that tunnel junctions have demonstrated efficient operation up to several thousand suns [8]. Moreover, when MJSCs operate inside an optical concentrator, spatial non-uniformity of the light impinging the solar cell aggravates series resistance losses [9]. To overcome these problems, the use of graphene in concentrator III-V MJSCs is proposed in this work. Graphene is incorporated covering both the front grid metallization and the topmost semiconductor layer to improve photocurrent extraction and therefore, the solar cell efficiency.

Since the first isolation of a single layer of graphene in 2004,[10] its unparalleled properties have allowed its integration in a wide variety of electro-optical applications (LEDs, solar cells, advanced sensors, etc.) [11-13]. In particular, in the field of solar cells, the use of graphene has been intensively studied in emerging architectures based on thin films, dye-sensitized, quantum dots, nanowires, perovskites, etc. resulting in efficiencies typically ranging 0.26-16.2% [14-20]. However, these efficiency figures are strongly affected by inner limitations in the efficiency of these solar cells. Therefore, the real potential of graphene -as an advanced energy material- to achieve very high efficient solar cells is still to be evinced.

Accordingly, in this paper we face the use of graphene in one of the most challenging types of solar cells, namely concentrator III-V multijunction solar cells. The defy of these solar cells derives from the following facts: a) this type of cells exhibits the highest power conversion efficiency[4-6], b) they are currently fabricated by means of a mature and robust technology at industrial level,[7] and c) their operation at concentration levels of 1,000 suns and beyond,[21]is an excellent benchmark to demonstrate the use of graphene to improve the photocurrent extraction and thus to increase the efficiency of concentrator III-V MJSCs. Although the experimental incorporation of graphene in this work has been implemented on lattice matched GaInP/Ga(In)As/Ge triple junction solar cells (3JSCs), the use of graphene we are proposing might be applied to the wide palette of III-V MJSC architectures available,[4, 6, 22, 23] as well as to other type of solar cells.

Through the paper, we firstly analyze the specific requisites and properties that graphene must fulfil for its successful integration in concentrator MJSCs. After that, we characterize the key optoelectronic properties of graphene layers to ensure their suitability in concentrator MJSCs. After that, we describe the experimental procedure to manufacture MJSCs with graphene. Finally, we compare the experimental performance of 3JSCs with and without graphene, which results in almost 1% (absolute) increase in the power conversion efficiency at around 1000 suns which increases as concentration raises. This represents a lower bound for the efficiency gain, therefore, future improvements that could be added to concentrator III-V MJSCs exhibiting efficiencies around 40% are discussed.

**Requisites for graphene in concentrator III-V MJSCs**

In this paper we use graphene on top of concentrator GaInP/Ga(In)As/Ge 3JSCs in conjunction with a conventional solar cell front metal grid (see Figure 1). The goal is that graphene creates paths of low electric resistance for the photocurrent extraction in parallel to those of the emitter, window and metal grid. These additional paths should result in a decrease of the series resistance which would in turn increase the fill factor ($FF$) and open circuit voltage ($V_{oc}$). Simultaneously, the optical transmission of graphene needs to be very high since otherwise the efficiency improvements achieved by the decrease in series resistance would be counterbalanced by the optical losses which would reduce the short circuit current density ($J_{sc}$).

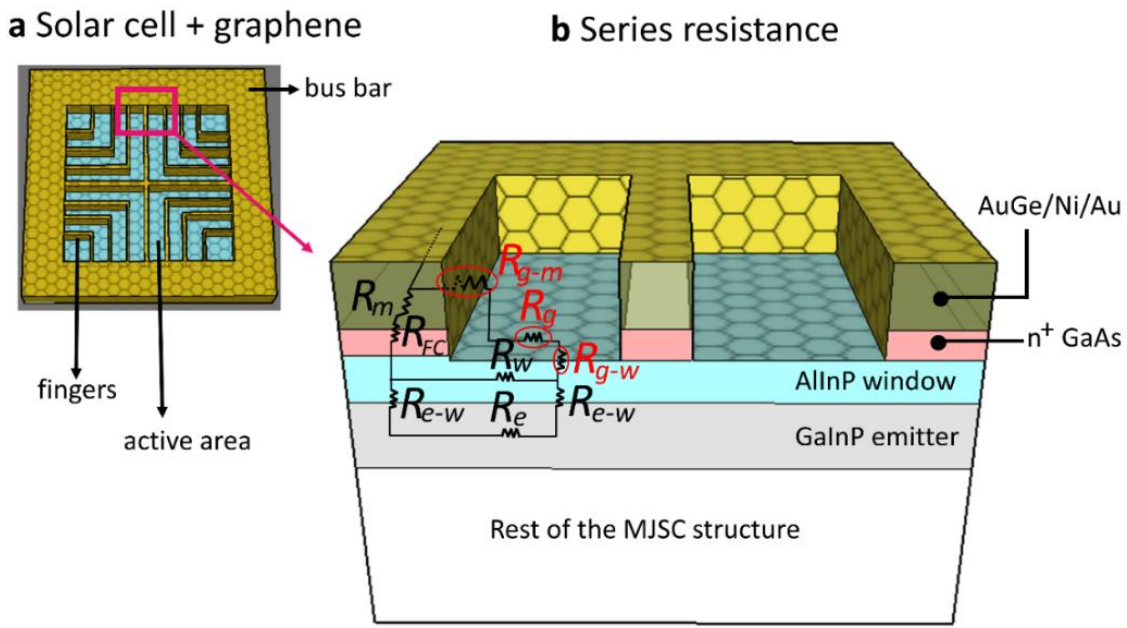

**Figure 1. Graphene integration in a concentrator III-V MJSC. Graphene is deposited in contact with both the front grid metallization and the semiconductor active area. a,** Graphene (black hexagons) on top of both the metallization (fingers + busbar, in yellow)

and semiconductor active area (AlInP window layer highlighted in cyan). **b**, Zoomed view of a small portion of Figure 1 a) where all the resistances participating in the front extraction of photocurrent have been included. Graphene integration provides alternative routes for current extraction marked as resistances encircled in red which represent: 1) the contact resistance between graphene and window layer ($R_{g-w}$); 2) the lateral graphene resistance ($R_g$) and 3) the contact resistance between graphene/metal grid ($R_{g-m}$).

Among the series resistance components of the 3JSC without graphene, the most important are schematized in Figure 1.b, namely:

- the sheet resistance of the front metal grid ($R_m$) with typical values of 0.05-0.20 $\Omega/\square$, which are good enough for concentrator solar cells with a typical size in the range of mm$^2$ required for a reduction, among others, of the series resistance and a low overheating at very high light concentrations;[1]

- the specific contact resistance between the front metal grid and the n$^+$-GaAs cap layer ($R_{FC}$) with values ranging $10^{-5}$-$10^{-6}$ $\Omega \cdot cm^2$,

- the top cell emitter sheet resistance ($R_{e\ top\ Sheet}$) with values of 300-500 $\Omega/\square$ (which includes both the AlInP window ($R_w$) and emitter sheet resistance ($R_e$)) derived from a trade-off of reducing the series resistance while keeping a high photogeneration.

When the same 3JSC incorporates graphene, three new paths for photocurrent extraction might participate, namely, $R_g$, $R_{g-w}$ and $R_{g-m}$ (encircled in red in Figure 1.b). The requisites for these resistances are: a) low sheet resistance of graphene to reduce the value of the lateral graphene resistance, $R_g$; b) low contact resistance between graphene and the front metal grid to minimize $R_{g-m}$, and c) low contact resistance between graphene and the top semiconductor (window layer) to get a low enough $R_{g-w}$.

The graphene sheet resistance, $R_g$, has to be as low as possible. In order to obtain the maximum benefit from the incorporation of graphene, a reasonable initial target is to aim for values equal or lower than the emitter sheet resistance, $R_{e\ top\ Sheet}$, i.e. 300-500 $\Omega/\square$. These values are achievable by Chemical Vapor Deposition (CVD)-grown graphene [24]. We have measured (see the Experimental section) a moderate mobility in the range of 750-2,000 $cm^2/V \cdot s$ in our graphene layers and a resistivity in the range of $2.4 \cdot 10^{-5}$-$5.7 \cdot 10^{-6}$ $\Omega \cdot cm$, resulting in a sheet resistance ranging from 200-500 $\Omega/\square$, which complies with the requirement of being in the range of $R_{e\ top\ Sheet}$. Further reduction in sheet resistance can be achieved by chemically doping graphene [25-27]. In relation to the graphene/front grid metallization interface, the requirement is to achieve an ohmic contact resulting in $R_{g-m}$ values around the specific front contact resistance between the front metal grid and the n$^+$-GaAs cap layer ($R_{FC}$). Figure 2.a shows the test structures manufactured to check the graphene/metal contact (in our 3JSCs the front metal grid is made of AuGe/Ni/Au) while Figure 2.b shows an example of the I-V curves measured, which confirm an ohmic behaviour. The results of the TLM (Transmission Line Method) analysis (detailed in the Experimental section) yield a specific contact resistance of the graphene/metal contact of $2 \cdot 10^{-4}$ $\Omega\ cm^2$, which is close to values reported in the literature [28].

a)

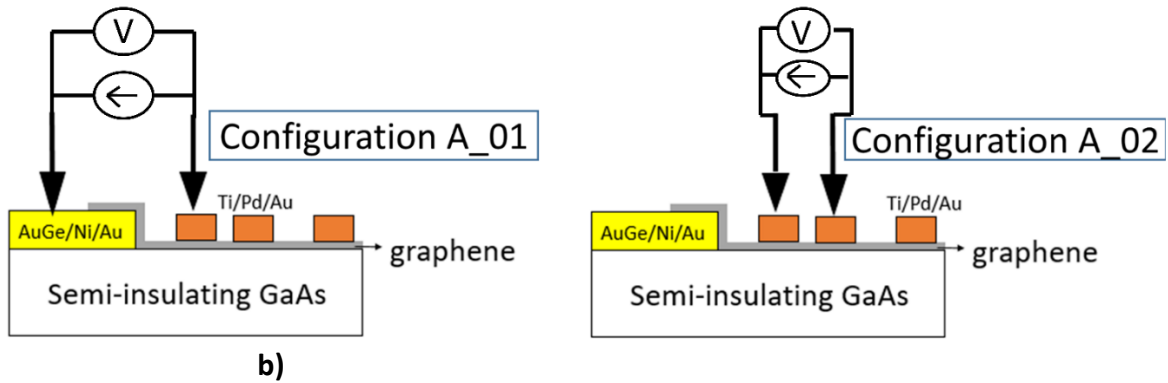

b)

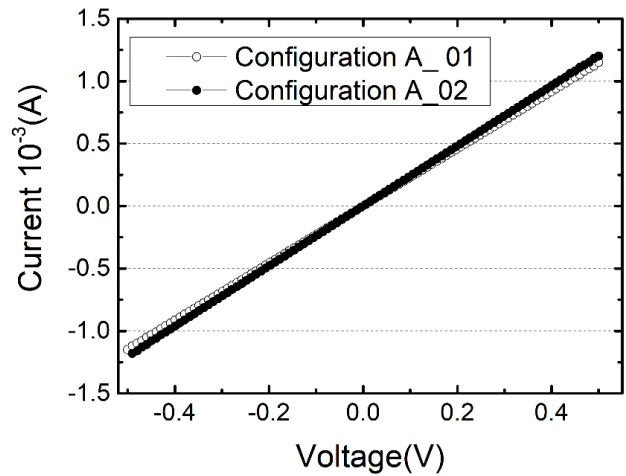

**Figure 2. Characterization of the graphene/metal contact. a**, Test structures used for the measurement of the contact resistance between graphene and the front metallization made of AuGe/Ni/Au and Ti/Pd/Au. Black arrows indicate the contacts used for current injection and voltage measurement. Configuration A_01 measures the I-V curve through AuGe/Ni/Au – graphene- Ti/Pd/Au while Configuration A_02 measures the I-V curve through Ti/Pd/Au- graphene- Ti/Pd/Au. **b**, Experimental I-V curves for both Configuration A_01 and Configuration A_02.

Regarding the equivalent resistance of the graphene/AlInP window layer interface, $R_{g-w}$, we firstly simulated its band alignment (Figure 3.a), which showed a high dependence on the AIInP doping level. Subsequently, we calculated the dark J-V curve of a 3JSC with an n-AlInP window layer of 40 nm with different doping levels and covered by a graphene layer with different work functions. Figure 3.b shows the simulated dark J-V curves taking into account the influence of both thermionic emission and tunneling mechanisms. Irrespective of AlInP doping level and for the three graphene work functions considered, tunneling appears as the dominant mechanism, its influence rising as the doping level increases. Figure 3.b shows that the impact of the graphene work function is very dependent upon the AlInP doping level. In fact, if the AlInP doping level increases over $10^{19}$ cm$^{-3}$, dark J-V curves are largely insensitive to variations in the graphene work function (which is affected by surface condition, i.e. contamination or oxidation of the AlInP window layer may vary the work function value) for the whole voltage range studied (i.e. the dashed black line, solid black line and squares virtually overlap in Figure 3.b). In such situation, current transport through the graphene/AlInP interface resembles an ohmic contact. Since the experimental doping level of the AlInP window layer of the 3JSCs in this study is in the range of $2 \cdot 10^{19}$ - $3 \cdot 10^{19}$ cm$^{-3}$, we may consider that our graphene/AlInP contact (assumed as ideal, i.e. without any interface contamination such as oxides) will behave as ohmic.

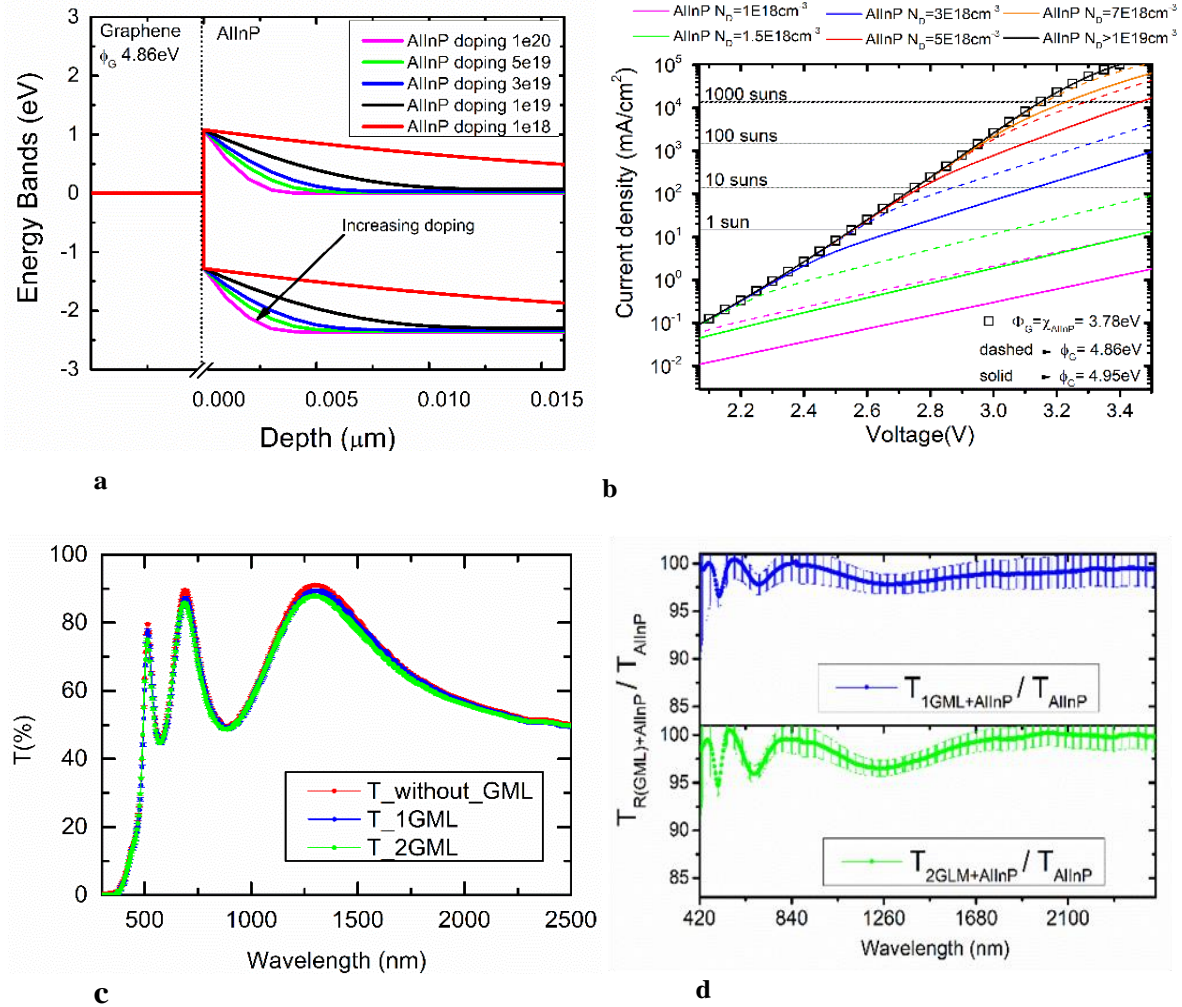

**Figure 3. Electrical and optical performance of the graphene/AlInP interface. a,** Simulated energy band alignment at equilibrium for a graphene monolayer (with a work function of 4.86 eV, which was experimentally measured) and an AlInP window layer as function of its doping level. **b,** Simulated dark J-V curves of a 3JSC integrating one graphene monolayer as a function of the graphene work function and the AlInP doping level. Simulations consider three values for the graphene work function to take into account the experimental values of our graphene measured by UPS, namely 4.86 eV (dashed) and 4.95 eV (solid) and also the ideal value of 3.78 eV (squares) coincident with the electronic affinity of AlInP. Different equivalent light concentration levels (1-1,000 suns) are included as guides to the eye by considering a $J_{sc} \sim 14$ mA/cm² at 1 sun. **c,**

Experimental Transmittance (T) of AlInP samples without graphene and with 1GML and 2GML.
**d,** Relative transmittance of 1GML and 2 GML deposited on AlInP samples (Eq 1). The uncertainty
in the optical transmittance has been assumed to be equal to the standard deviation

Further reductions on the series resistance are possible by stacking several graphene
layers on top of each other. Ideally, the larger the number of graphene layers, the lower the
lateral series resistance but the lower the optical transmission too. In this way, a trade-off
arises around the optimum number of graphene layers, which will be discussed below.

Regarding the optical behaviour, we have studied the optical impact of transferring
graphene on AlInP window layers by measuring the spectral reflectance ($R$) and
transmittance ($T$), as indicated in the Experimental Section. In this work, we have used
graphene grown by CVD, as this growth technique offers excellent control of the material
properties over large areas and it is scalable to industrial levels [29, 30].

Figure 3.c shows the experimental transmittance of the AlInP samples without
graphene and with one and two graphene monolayers (1GML and 2GML). The simulated
transmittance curves by using the Generalized Transfer Matrix Method,[31] with the
refractive index and extinction coefficients determined for our CVD graphene [32] (not
shown in Figure 3.c for the sake of clarity), fit perfectly the experimental ones for the AlInP
structures with and without graphene. As expected, transmittance decreases as graphene
thickness increases for all wavelengths.

The relative transmission of graphene layers ($T_{R,GML}(\lambda)$) can be obtained by using
the measured transmittance of the AlInP sample ($T_{AlInP}(\lambda)$) together with the transmittance
of the AlInP samples covered with 1GML and 2GML ($T_{GML+AlInP}(\lambda)$):

$$T_{R,GML}(\lambda) = \frac{T_{GML+AlInP}(\lambda)}{T_{AlInP}(\lambda)} \qquad\qquad (1)$$

Figure 3.d depicts these results. The resulting average transmittance for 1GML and 2GML within the 400-1700 nm range (which is that of interest for the kind of 3JSCs of this work) is, 98.7% and 98.0%, respectively. From the transmittance curves of Figure 3.d and the measured reflectance curves (not shown), we have calculated the corresponding absorbance for 1GML and 2GML whose average values are 1.2 and 3.7%, respectively for the visible range (400-700 nm). These values are consistent with that of,[33] who predicted by band theory an absorbance of 1% for graphene on silicon substrates, which have a refractive index of 3.4, close to that of AlInP. It should be noticed, that this is a different case from that of 2.3% absorbance of suspended graphene[34].

**Experimental**

Integration of CVD graphene on 3JSCs involves several steps that will be sequentially described through this experimental section. These steps start from the growth of both graphene monolayers (by CVD) and III-V MJSCs (by Metal Organic Vapor Phase Epitaxy, MOVPE) up to the graphene transfer and isolation of each individual solar cell device. Additionally, electro-optical measurements of CVD graphene monolayers prior to their integration on 3JSCs as well as characterization of complete solar cells with and without graphene are described.

*CVD graphene growth*

Graphene monolayers were synthesized using CVD. Graphene was grown following the recipes of Kong's group at MIT [35]. Copper (Cu) foils (~25 µm thick, 99.9%) were used as the catalytic substrates. A conventional furnace was used for the graphene growth at temperatures of around 1000°C using a mixture of methane ($CH_4$) and hydrogen ($H_2$) as precursor gases. Cu foils of ~2.5 cm × 10 cm were pre-treated by dipping them into a Cu etchant. This step reduces native oxides on the Cu surface that could affect graphene growth [36]. Further details on the CVD growth can be found elsewhere [29, 30, 35, 37].

*Graphene wet transfer for its characterization*

Graphene wet transfer was also carried out following the procedure from Kong´s group [38]. In the wet transfer method, a polymer film is deposited on the graphene/Cu stack and then the sample is baked at 130ºC for 10 min. Immediately after, the catalyst metal is etched away with a Cu etchant. The resulting PMMA/graphene stack is cleaned in HCl:$H_2O$. These steps help remove some debris from the Cu etchant such as oxidized metal particles. The PMMA/graphene film is then scooped, deposited on the substrate of interest and dried with $N_2$. Afterwards, baking steps are carried to promote graphene adhesion on the semiconductor and then PMMA is removed by soaking the sample in acetone. This step sequence does not ensure a completely clean surface since, in the end, graphene has been exposed to 1) the metal etchant, 2) polymer residues and 3) water and dust particles that could adhere during the different transferring steps and eventually strongly affect the final properties of the material [25, 26]. Therefore, a final thermal treatment is used to get rid of such contaminants.

*Electro-optical characterization of CVD graphene*

Transmission Line Method (TLM) and Van der Pauw measurements were carried out on CVD graphene monolayers for their electrical characterization (graphene sheet resistance, mobility and doping levels among others). Graphene monolayers were transferred by the wet transfer method into $SiO_2$/Si substrates. TLM patterns were fabricated onto 1GML/$SiO_2$/Si by using e-gun evaporation of 5.5 nm Ti/45 nm Pd/110 nm Au. Conventional photolithographic techniques were used to define pads of an area of 400 μm x 150 μm with pads distances of 90/180/360/720/1440 μm. In order to get rid of the lateral deviation of the current a second photolithography step for TLM isolation was needed. A dry etching was applied to etch graphene from the desired areas.

Van der Pauw measurements were performed after the graphene transfer to $SiO_2$/Si substrates using a homemade set-up. For simplicity, no photolithography was applied. Instead, a silver paint was used to form an ohmic contact pad by applying it on 4 edges of a square shaped graphene layer. By applying a current of 1 mA and a magnetic field of 2900 Gauss, the Van der Pauw technique results in the determination of the graphene sheet resistance, mobility and carrier density.

Graphene work function in contact with the AlInP window layer was obtained using Ultraviolet Photoelectron Spectroscopy (UPS). A photon source of He (I), 21.21 eV, was used. Since work function is affected by the surface condition, i.e. contamination or oxidation may have a significant influence on the work function value obtained, a previous Argon (Ar) cleaning treatment was performed for some samples to see that influence. This cleaning treatment consisted in an $Ar^+$ sputtering with an energy of ~1 keV for 60 seconds.

Graphene/AlInP work function with and without cleaning treatment resulted in 4.95 eV and 4.86 eV, respectively.

Spectral transmittance of the topmost part of the 3JSCs cells was determined by measuring the stack consisted of graphene (1GML and 2GML) transferred onto an AlInP window layer stack with a transparent epoxy on top of a glass holder. Prior to graphene transfer, transmittance of AlInP window layer on the glass holder was measured. Therefore, the relative transmittance of graphene ($T_{R,GML}(\lambda)$) was obtained by using the transmittance of the AlInP window layer on top of a glass holder for its normalization following Equation 1. Spectral transmittance measurements from visible to middle infrared range were done by means of a Perkin Elmer lambda 1050 Spectrometer.

*Manufacturing of concentrator triple junction solar cells (3JSCs)*

Lattice matched GaInP/Ga(In)As/Ge triple junction epiwafers with energy bandgaps of 1.89 eV, 1.41 eV and 0.6 eV, respectively, were grown on p-type gallium doped Ge (100) wafers with 6° misorientation to [111] by MOVPE. $AsH_3$, $PH_3$, TMGa, TMAl, TMIn, DETe, DMZn, $DTBSi_2$ and $CBr_4$ were used as constituent and dopant precursors. Details on the semiconductor structure and the epitaxial growth can be found elsewhere [39-43]. Epiwafers were processed following conventional photolithographic techniques. The front grid geometry used was the inverted square configuration with an active area of 1 mm$^2$ [44]. AuGe/Ni/Au is deposited on the front n-GaAs cap layer which is subsequently etched in regions uncovered by metals, leaving the AlInP window layer exposed. Pure gold was deposited on the backside to form the rear contact on the Germanium substrate. Both contacts (top and bottom) were annealed in forming gas at 375°C for 180 s and 200°C for 120 s, respectively. No antireflection coatings (ARCs) were deposited on the solar cells in order to ease the comparison between solar cells with and without graphene.

*Integration of graphene on concentrator 3JSCs*

Graphene monolayers synthesized by CVD were transferred by the already described wet transfer method on triple junction solar cells. In order to achieve a fair comparison between 3JSCs with and without graphene, CVD-graphene was deposited only on one half of the epiwafer so, the epistructure and device processing were identical for all solar cells with and without graphene. Additionally, when a second monolayer (2GML) of graphene was deposited, a subsequent wet transfer process was repeated. Although thermal treatments between 300-350ºC are desired as a last step of the graphene transfer process, during this work the presence of both front and back metallization on the solar cell wafer limited this temperature down to 200ºC to avoid their deterioration.

*Singulation of 3JSCs incorporating graphene (RIE + mesa etching + dicing)*

One of the last steps of the solar cell manufacturing is the electrical isolation of each individual device in the wafer by mesa etching. Since graphene is chemically inert to wet etchants and is placed covering the semiconductor layers (see Figure 4), a purely chemical mesa etching will not be effective if graphene is not previously removed from the mesa trenches. Thus, prior to mesa isolation, a second photolithography step was required to protect graphene at areas where it has to be preserved (i.e. on the active area and on the front grid metallization). Graphene was etched away from the isolation trenches via RIE and afterwards the mesa etching was carried out. Finally, the epiwafers containing the solar cells were rinsed in acetone and methanol to remove the photoresist. After that, solar cells with and without graphene were diced and individually packaged onto printed circuit boards (PCBs). No ARC was deposited on the devices to facilitate the comparison between solar cells with and without graphene as well as to determine more clearly the graphene optical impact on 3JSCs. Figure 4 visually summarizes the key steps in the manufacturing of 3JSCs with graphene.

*Characterization of graphene-based 3JSCs*

The characterization of 3JSCs with and without graphene consisted of the dark I-V, external quantum efficiency (EQE), concentrator response and electroluminescence (EL) mapping. I-V curves were measured with the four point probe technique to supress the influence of the series resistance of the probe and wires on the measurements. The setup employed for the EQE consists of a Xe lamp used as white light source, which passes through a Horiba Jobin Yvon monochromator (TRIAX180) and a filter wheel. Further details on the system and the measurement can be found in [45]. Measurements under concentration were performed with a flash simulator. For a detailed explanation of the concentration set-up, the reader is referred to [46]. Regarding EL measurement, driving the cells into forward bias makes them to behave as Light Emitting Diodes (LEDs). A current injection of 100 mA was used to emulate the solar cell current photogeneration at concentration levels around 900×, (assuming $J_{SC}^{1\times} \sim$ 10-11 mA/cm$^2$ without ARC). The emitted light is recorded with a CCD camera (CCi4 MV1-D1312I Photon Focus) in a home-made EL set-up. The electroluminescence images investigated in this work, were those of the GaInP top subcell only (emission at 650-690 nm), for which a short pass filter is used (cut off wavelength 800 nm).

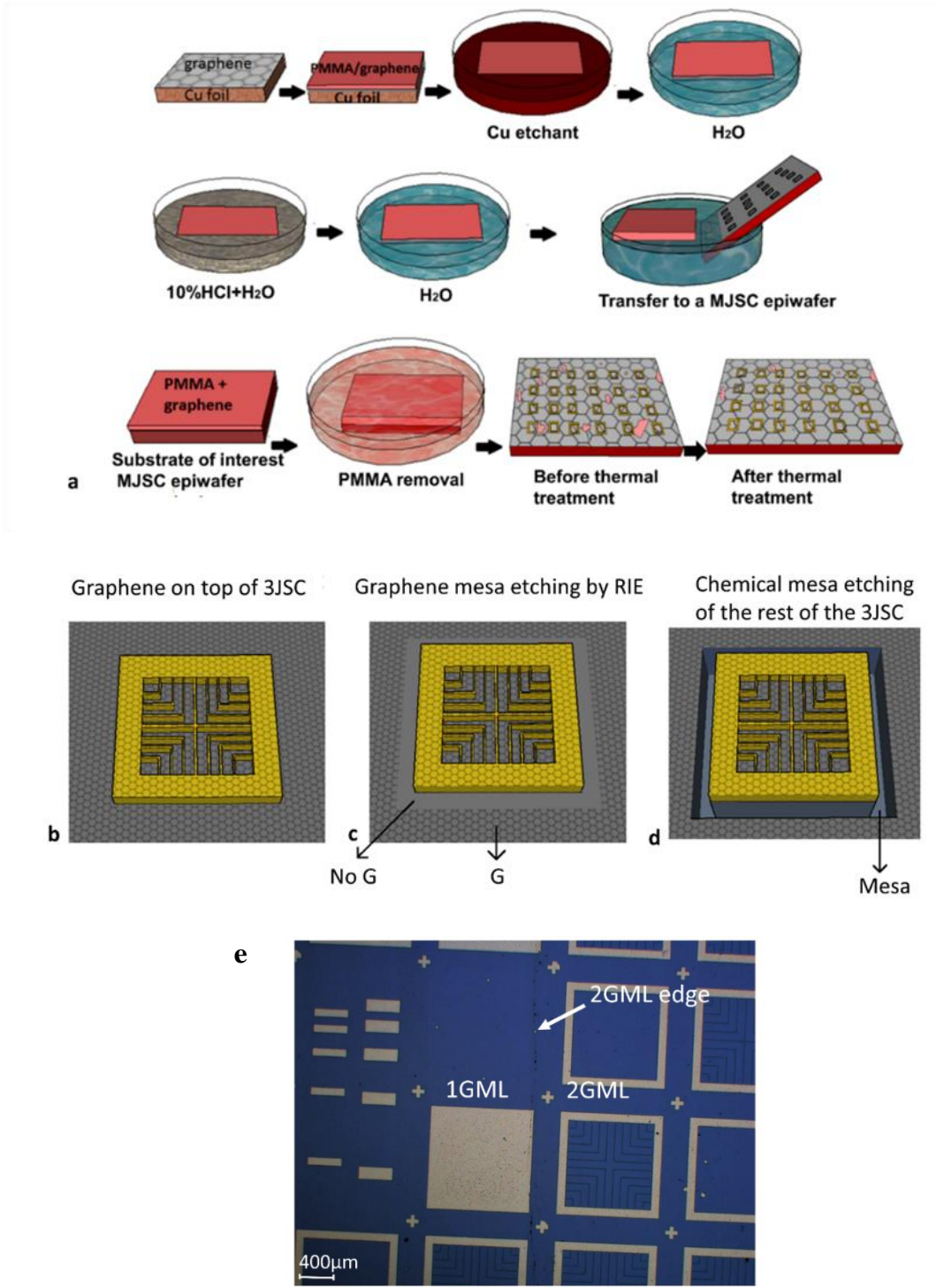

**Figure 4 Manufacturing stages of III-V MJSCs incorporating graphene. a,** Steps of graphene wet transfer method for its integration on a III-V MJSC epiwafer. **b-d**, Isolation of each individual solar cell on the epiwafer. (b), Graphene on top of a solar cell (c), Graphene RIE on top of the mesa area (d), Final mesa after the wet chemical etching until the Ge substrate is reached. **e**, Optical microscope image of one (1GML) and two graphene monolayers (2GML) transferred on top of some solar cells where the edge of the second graphene layer provides the identification between 1GML and 2GML regions.

**Results and Discussion**

In order to characterize the solar cells with and without graphene, we firstly checked that graphene layer conformally follows the solar cell surface by means of Atomic Force Microscopy (AFM) in tapping mode. As Figure 5.a left shows, the adherence of graphene between metal fingers and the solar cell active region is reasonable, and the only regions where the graphene layer does not touch the active region is next to the metal fingers. In the case of the busbar (Figure 5.a. right), the abrupt step denotes a better adherence between graphene and active area. Therefore, we can expect that a significant part of the solar cell surface is in intimate contact with the graphene layer. This assumption is confirmed by EL maps measurements of 3JSCs without graphene and with 1GML (Figure 5.b) which show similar performance indicating that the solar cell with 1GML has not regions with poor contact between the graphene layer and the solar cell surface. A more careful examination of Figure 5.b shows that the cell with 1GML exhibits at the center a higher EL emission (less cyan color) than the cell without graphene (as Figure 5.c highlights), which anticipates a better current distribution thanks to graphene.

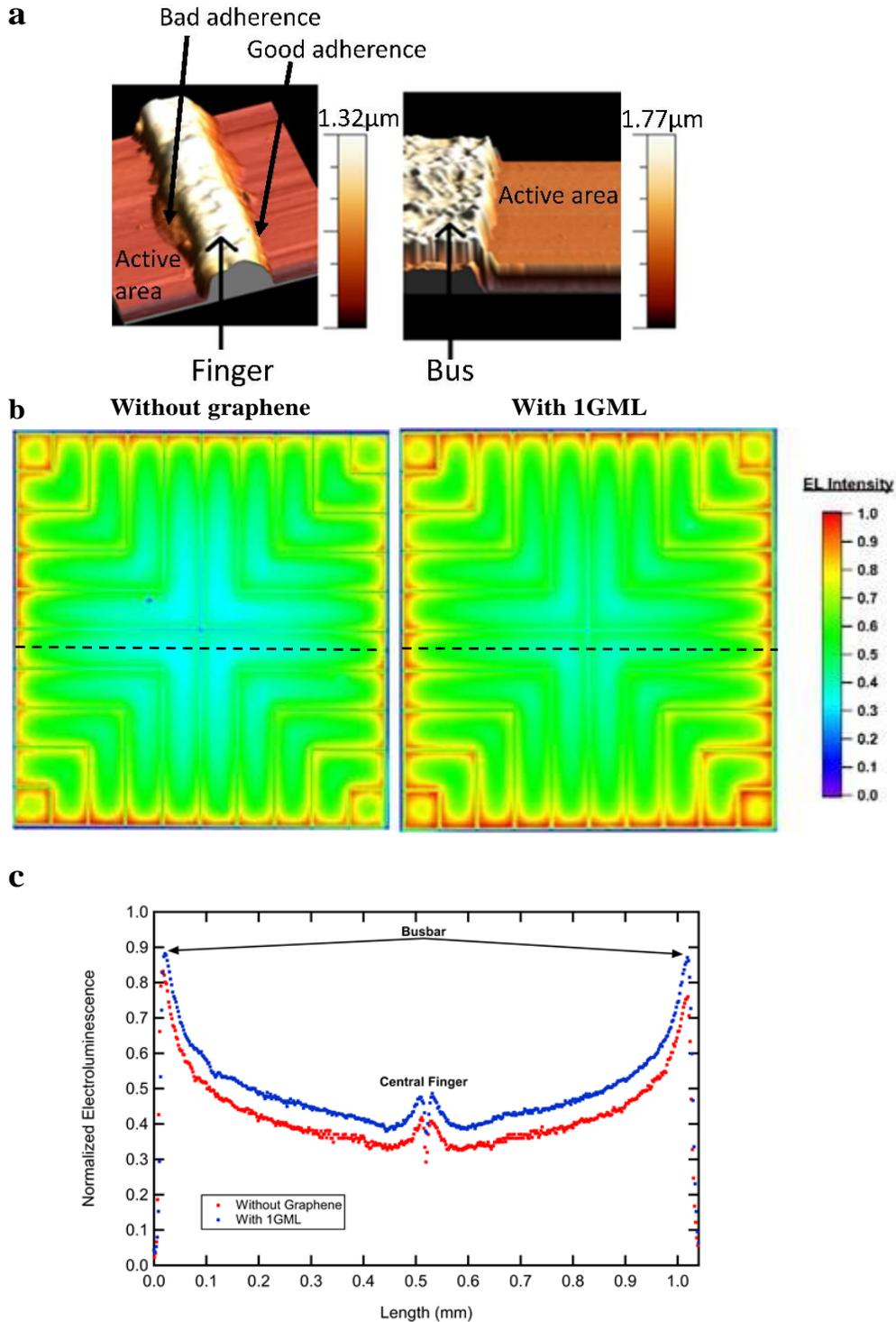

**Figure 5. Characterization of the graphene/solar cell surface. a**, AFM topography with the z-scale scan of the front metallization of a 3JSC incorporating one graphene monolayer.

Active area and a portion of a 3.5 µm width metal finger both covered with graphene (left).

Busbar and active area covered with graphene (right). **b,** Experimental false color maps of normalized electroluminescence of the GaInP top subcell of a 3JSC with 1GML and without graphene. The solar cell area inside busbar is 1 mm$^2$ with the geometry shown in Figure 4.b-d. Electroluminescence at a forward current of 100 mA shows emission at the metal fingers than can be attributed to a Lambertian emission detected by the EL camera. Metal busbars are not shown**. c**, Comparison between the EL across the black dotted lines of Figure 5.b for cells with 1GML and without graphene.

The EQE is presented in Figure 6.a for the GaInP top, Ga(In)As middle and Ge bottom subcells of the 3JSCs with and without graphene. As it can be observed in Figure 6.a, a very small decrease in EQE for some wavelengths is detected after graphene integration. To quantify such decrease, the short circuit current density ($J_{sc}$) was calculated for several solar cells from the convolution of their EQE with the reference spectrum for concentration (AM1.5d ASTM G173-03) [47]. That results in a relative $J_{sc}$ decrease with respect to solar cells without graphene in the 0-1.8% range for 1GML while 2.0-3.8% for 2GML, which are consistent with the average absorbance of 1.2% and 3.7% for 1GML and 2GML, respectively, described above. Concentration measurements at different irradiance levels were performed on the same solar cells without ARC in order to ease the comparison between solar cells with and without graphene. Figures 6.b-d show the evolution of $FF$, $V_{oc}$ and power conversion efficiency ($\eta$) as a function of concentration. Figure 6.b exhibits $FF$ values similar for solar cells with and without graphene at low concentrations, thus being the influence of graphene almost negligible. It is not until medium concentrations ($\geq 300\times$) when $FF$ of solar cells with graphene is higher. In fact, $FF$ improvement increases with concentration, reaching an increment of around 4% (absolute) at approximately 1,000×. This can be explained by the decrease in the series resistance caused by graphene, which minimizes the ohmic losses during the photocurrent extraction, with the subsequent improvement in *FF*. The *FF* data fitting to our solar cell models shows a series resistance of 34 m$\Omega$·cm$^2$ for the 3JSCs without graphene which is reduced to 22 m$\Omega$·cm$^2$ for the cells with 1GML (i.e. a 35% relative reduction of series resistance). The addition of a second graphene monolayer results in an additional *FF* improvement. Nevertheless, this second improvement is lower than expected for 1GML+1GML [24]. A possible explanation for this fact could be rely on the effect of contaminants (PMMA, water molecules, etc.) from the wet transfer process that might have been trapped between the two graphene monolayers stack (as was reported in previous works),[32] hindering photocurrent transfer between them.

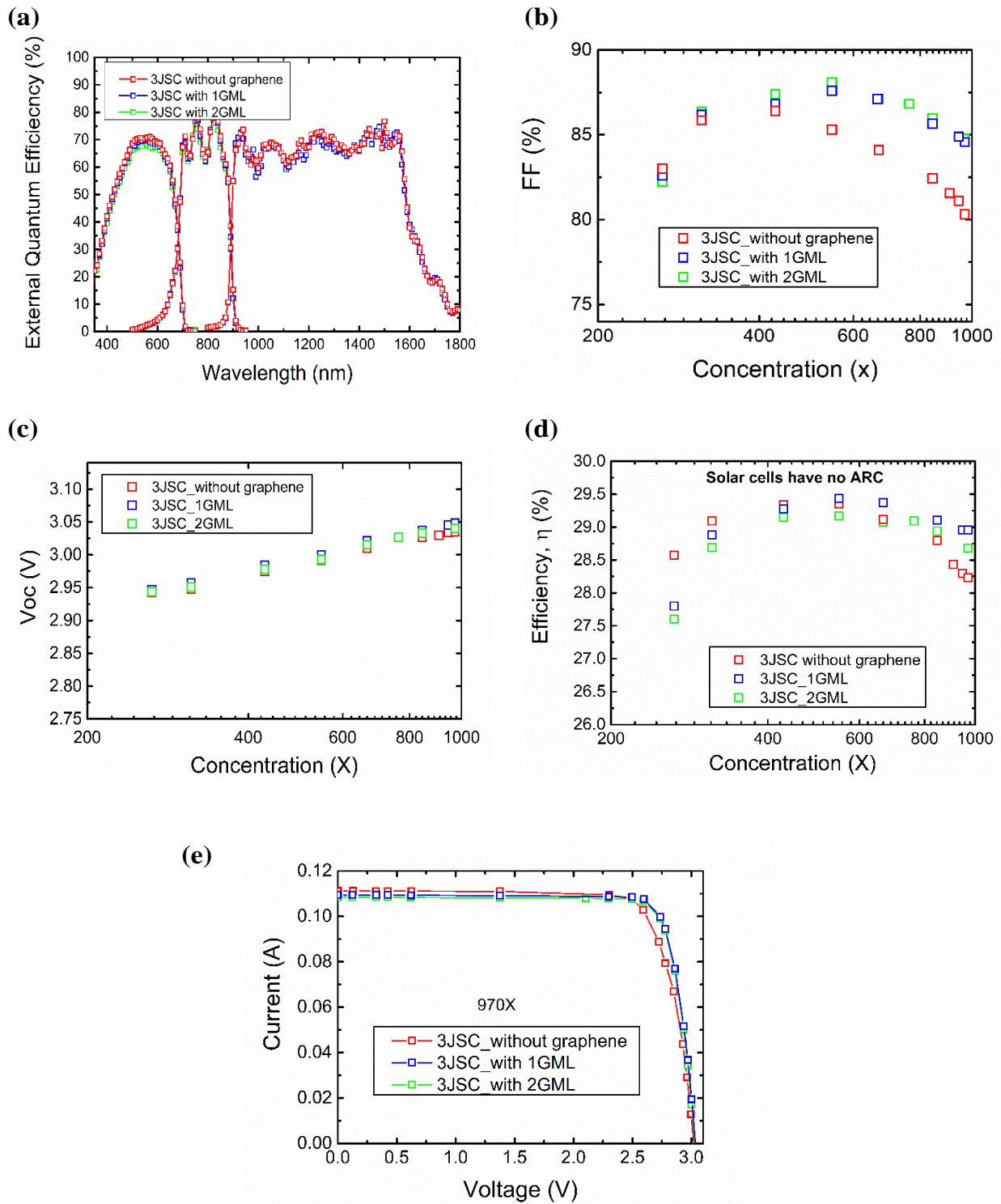

Figure 6. Characterization of 3JSCs with and without graphene. None of the 3JSCs have

ARC so, assuming a typical efficiency increase of 35% due to ARC, these solar cells with

**ARC would result in absolute efficiencies higher than 39% at 1,000 suns. Each point of all the figures is the average of three different solar cells. a**, EQE of the GaInP, Ga(In)As and Ge subcells of a 3JSC without graphene and with 1GML and 2MLG. The Ge subcell with 2MLG is not shown. **b-e**, Performance as a function of light concentration of 3JSCs without and with graphene (with 1GML and 2GML). (b), Fill factor (*FF*). (c), Open circuit-voltage (*V$_{oc}$*). (d)**,** Power conversion efficiency (*η*). (e), I-V curves at the highest concentration in each 3JSC design. Solar cells have no ARC.

Figure 6.c shows the evolution of *V$_{oc}$* vs. concentration for the same 3JSCs with 1GML, 2GML and without graphene showing a very similar performance. Figure 6.d shows that efficiency of 3JSCs without graphene is higher at low concentrations but is outdone at high irradiance levels by 3JSCs with graphene. The difference at high concentrations shows up clearly in the I-V curves of Figure 6.e. This is because *FF* and *V$_{oc}$* improve with the use of graphene thanks to the series resistance reduction, but at the same time, *J$_{sc}$* slightly decreases because of graphene optical absorption. This trade-off leans towards *FF* and *V$_{oc}$* only for high enough concentrations. As an example, at 1,000× absolute efficiency gains close to 1% (1GML) and 0.5% (2GML) are achieved, with respect to 3JSCs without graphene and these efficiency gains are even higher at concentrations beyond 1000 suns. This fact can be explained in terms of the transparency/resistance trade-off: for 2GML the decrease in resistance obtained in our experiment does not improve the *FF* so much as to counterbalance the higher optical losses. Of course, this could change if better photocurrent transfer would have been achieved between the two graphene monolayers but until this problem is not experimentally solved, our work shows that the optimum number

of graphene layers is one. It is worth mentioning that none of the 3JSCs have ARC so, assuming a typical power conversion efficiency increase of 35% due to ARC, these solar cells with ARC would result in absolute efficiencies higher than 39% at 1,000 suns.

To further improve the results of this paper, the deposition of graphene on top of the window layer could be done prior to the evaporation of the front grid metallization. In this way, graphene will be deposited directly on a completely planar and clean surface without the metal fingers. This would reduce both the resistance of the graphene/AlInP window layer interface, $R_{g-w}$, and the resistance of the graphene/front grid metallization interface, resulting in a higher $FF$. Further reductions in graphene sheet resistance can be achieved by chemical doping. Besides, this alternative approach of integrating graphene prior to the front grid metallization would allow a higher flexibility in the choice of the annealing time and temperature of graphene for a more effective suppression of surface residue. This would result in a reduction of the optical absorption of graphene with the subsequent increase of $J_{sc}$. Therefore, these increases of $FF$ and $J_{sc}$ will produce an improvement in the power conversion efficiency of concentrator III-V MJSCs.

**Summary and Conclusions**

We have assessed CVD-graphene incorporation atop concentrator III-V MJSCs. Firstly, the requisites and properties for graphene to be integrated in those cells were stated and then, a wide optoelectronic characterization of graphene layers was carried out in order to confirm that our CVD graphene complies with such requirements.

High-concentrator GaInP/Ga(In)As/Ge triple-junction solar cells were manufactured on epiwafers grown by MOVPE. Graphene monolayers were transferred onto one half of the 3JSC structures by using wet transfer method. In this way, a fair comparison between 3JSCs with and without graphene can be done, since their epistructures and device processing were identical. No ARC were deposited in order to ease that comparison.

Graphene incorporation onto triple-junction solar cells shows an effective decrease of the series resistance (35% relative), leading to an increase in *FF* of 4% (absolute) at concentrations of 1,000×. EL measurements confirm a better spreading of the photocurrent in solar cells with graphene. Simultaneously, the optical absorption of graphene produces a relative $J_{sc}$ decrease in the range of 0-1.8%. As a result, an absolute efficiency improvement close to 1% at concentrations of 1,000× was achieved with respect to 3JSCs without graphene and this improvement increases with concentration. The impact of one and two graphene monolayers was also evaluated showing that a single graphene layer is better than two. These results highlight the potential of graphene to reduce the impact of series resistance in concentrator MJSCs and pave the way to the exploitation of CPV designs working over a thousand suns.

## Acknowledgements

This work has been supported by the Spanish MINECO through the projects TEC2017-83447-P, PCIN-2015-181-C02-01 and PCIN-2015-181-C02-02 and by the Comunidad de Madrid through the project TEFLON-CM (Y2018/EMT-4892) which is also funded with FEDER funds. Iván García is funded by the Spanish Programa Estatal de Promoción del Talento y su Empleabilidad through a Ramón y Cajal grant (RYC-2014-15621). Iván Lombardero acknowledges the financial support from the Spanish Ministerio de Educación, Cultura y Deporte through the Formación del Profesorado Universitario grant with reference FPU14/05272. Tomás Palacios acknowledges support from the NSF Grant No. DMR 1231319 and the AFOSR Grant No. FA9550-15-1-0514.

**References**

[1]     C. Algora,"Very-high-concentration challenges of III-V multijunction solar cells", in *Concentrator Photovoltaics*, Springer: Heidelberg, Germany, 2007.

[2]     C. Algora, and I. Rey-Stolle,"The interest and potential of ultra-high concentration", in *Next Generation of Photovoltaics*, Springer, 2012.

[3]     P. Pérez-Higueras, and E.F. Fernández, "High concentrator photovoltaics: fundamentals, engineering and power plants*"*, Springer, 2015.

[4]     F. Dimroth, M. Grave, P. Beutel, et al., "Wafer bonded four‐junction GaInP/GaAs//GaInAsP/GaInAs concentrator solar cells with 44.7% efficiency". *Prog Photovoltaics*, 22, pp 277-282, 2014.

[5]     R.M. France, J.F. Geisz, I. García, et al., "Quadruple-junction inverted metamorphic concentrator devices". *IEEE J. Photovoltaics*, 5, pp 432-437, 2015.

[6]     V. Sabnis, H. Yuen, and M. Wiemer, "High-efficiency multijunction solar cells employing dilute nitrides*"*. *AIP Conference Proceedings*. 2012.

[7]     C. Algora, and I. Rey-Stolle, "Handbook of concentrator photovoltaic technology*"*, John Wiley & Sons, 2016.

[8]     E. Barrigón, I. García, L. Barrutia, et al., "Highly conductive p++‐AlGaAs/n++‐GaInP tunnel junctions for ultra‐high concentrator solar cells". *Prog Photovoltaics*, 22, pp 399-404, 2014.

[9]     R. Herrero, M. Victoria, C. Domínguez, et al., "Concentration photovoltaic optical system irradiance distribution measurements and its effect on multi‐junction solar cells". *Prog Photovoltaics*, 20, pp 423-430, 2012.

[10]    K.S. Novoselov, A.K. Geim, S.V. Morozov, et al., "Electric field effect in atomically thin carbon films". *Science*, 306, pp 666-669, 2004.

[11]    C.-Y. Cho, M. Choe, S.-J. Lee, et al., "Near-ultraviolet light-emitting diodes with transparent conducting layer of gold-doped multi-layer graphene". *J. Appl. Phys*, 113, 113102, pp, 2013.

[12]    N. Yang, J. Zhai, D. Wang, et al., "Two-dimensional graphene bridges enhanced photoinduced charge transport in dye-sensitized solar cells". *ACS Nano*, 4, pp 887-894, 2010.

[13]    J. Liu, Y. Xue, M. Zhang, et al., "Graphene-based materials for energy applications". *MRS Bulletin*, 37, pp 1265-1272, 2012.

[14]    C.X. Guo, H.B. Yang, Z.M. Sheng, et al., "Layered graphene/quantum dots for photovoltaic devices". *Angew Chem*, 49, pp 3014-3017, 2010.

[15]     S. Hou, X. Cai, H. Wu, et al., "Nitrogen-doped graphene for dye-sensitized solar cells and the role of nitrogen states in triiodide reduction". *Energ Environ Sci*, 6, pp 3356-3362, 2013.

[16]     A.M. Munshi, D.L. Dheeraj, V.T. Fauske, et al., "Vertically aligned GaAs nanowires on graphite and few-layer graphene: generic model and epitaxial growth". *Nano Lett*, 12, pp 4570-4576, 2012.

[17]     Z. Zhu, J. Ma, Z. Wang, et al., "Efficiency enhancement of perovskite solar cells through fast electron extraction: the role of graphene quantum dots". *JACS*, 136, pp 3760-3763, 2014.

[18]     F. Biccari, F. Gabelloni, E. Burzi, et al., "Graphene - Based Electron Transport Layers in Perovskite Solar Cells: A Step - Up for an Efficient Carrier Collection". *Adv. Energy Mater*, 7, 1701949, pp, 2017.

[19]     C.X. Guo, G.H. Guai, and C.M. Li, "Graphene based materials: enhancing solar energy harvesting". *Adv. Energy Mater*, 1, pp 448-452, 2011.

[20]     Z. Yin, J. Zhu, Q. He, et al., "Graphene - based materials for solar cell applications". *Adv. Energy Mater*, 4, 1300574, pp, 2014.

[21]     M. Ochoa, E. Barrigón, L. Barrutia, et al., "Limiting factors on the semiconductor structure of III – V multijunction solar cells for ultra - high concentration (1000 – 5000 suns)". *Prog Photovoltaics*, 24, pp 1332-1345, 2016.

[22]     J.F. Geisz, A. Duda, R.M. France, et al., "Optimization of 3-junction inverted metamorphic solar cells for high-temperature and high-concentration operation". *AIP Conference Proceedings*. 2012.

[23]     D. Derkacs, R. Jones-Albertus, F. Suarez, et al., "Lattice-matched multijunction solar cells employing a 1 eV GaInNAsSb bottom cell". *J. Photonics Energy*, 2, 021805, pp, 2012.

[24]     F. Bonaccorso, Z. Sun, T. Hasan, et al., "Graphene photonics and optoelectronics". *Nat. Photonics*, 4, 611, pp, 2010.

[25]     A. Capasso, M. De Francesco, E. Leoni, et al., "Cyclododecane as support material for clean and facile transfer of large-area few-layer graphene". *Appl. Phys. Lett*, 105, 113101, pp, 2014.

[26]     J. Leclercq, and P. Sveshtarov, "The Transfer of Graphene: A Review". *Bulg. J. Phys*, 43, 121, pp, 2016.

[27]     P. Avouris, and C. Dimitrakopoulos, "Graphene: synthesis and applications". *Mater. Today*, 15, pp 86-97, 2012.

[28]     T. Palacios, A. Hsu, and H. Wang, "Applications of graphene devices in RF communications". *IEEE. Commun. Mag*, 48, pp, 2010.

[29]     P.R. Somani, S.P. Somani, and M. Umeno, "Planer nano-graphenes from camphor by CVD". *Chem. Phys. Lett.*, 430, pp 56-59, 2006.

[30]     X. Li, W. Cai, J. An, et al., "Large-area synthesis of high-quality and uniform graphene films on copper foils". *Science*, 324, pp 1312-1314, 2009.

[31]     E. Centurioni, "Generalized matrix method for calculation of internal light energy flux in mixed coherent and incoherent multilayers". *Appl Opt*, 44, pp 7532-7539, 2005.

[32]     E. Ochoa-Martínez, M. Gabás, L. Barrutia, et al., "Determination of a refractive index and an extinction coefficient of standard production of CVD-graphene". *Nanoscale*, 7, pp 1491-1500, 2015.

[33]   C. Lee, J.Y. Kim, S. Bae, et al., "Optical response of large scale single layer graphene". *Appl. Phys. Lett*, 98, 071905, pp, 2011.

[34]   R.R. Nair, P. Blake, A.N. Grigorenko, et al., "Fine structure constant defines visual transparency of graphene". *Science*, 320, pp 1308-1308, 2008.

[35]   A. Reina, X. Jia, J. Ho, et al., "Large area, few-layer graphene films on arbitrary substrates by chemical vapor deposition". *Nano Lett*, 9, pp 30-35, 2008.

[36]   S.M. Kim, A. Hsu, Y.-H. Lee, et al., "The effect of copper pre-cleaning on graphene synthesis". *Nanotechnology*, 24, 365602, pp, 2013.

[37]   W. Fang, A.L. Hsu, Y. Song, et al., "Asymmetric growth of bilayer graphene on copper enclosures using low-pressure chemical vapor deposition". *ACS Nano*, 8, pp 6491-6499, 2014.

[38]   G.B. Barin, Y. Song, I. de Fátima Gimenez, et al., "Optimized graphene transfer: Influence of polymethylmethacrylate (PMMA) layer concentration and baking time on graphene final performance". *Carbon*, 84, pp 82-90, 2015.

[39]   B. Galiana, E. Barrigón, I. Rey-Stolle, et al., "Compositional analysis and evolution of defects formed on GaInP epilayers grown on Germanium". *Superlattices Microstruct*, 45, pp 277-284, 2009.

[40]   E. Barrigón Montañés, "Development of GaInP/GaInAs/Ge TRIPLE-junction solar cells for CPV applications", Universidad Politécnica de Madrid, 2014**.**

[41]   L. Barrutia Poncela, "Optimization pathways to improve GaInP/GaInAs/Ge triple junction solar cells for CPV applications", Universidad Politécnica de Madrid, 2017**.**

[42]   W. He, S. Lu, J. Dong, et al., "Structural and optical properties of GaInP grown on germanium by metal-organic chemical vapor deposition". *Appl. Phys. Lett*, 97, pp, 2010.

[43]   E. Barrigón, I. Rey-Stolle, B. Galiana, et al., "GaInP/GaInAs/Ge triple junction solar cells for ultra high concentration*"*. *CDE*. 2009. IEEE.

[44]   C. Algora, and V. Díaz, "Influence of series resistance on guidelines for manufacture of concentrator p‑on‑n GaAs solar cells". *Prog Photovoltaics*, 8, pp 211-225, 2000.

[45]   E. Barrigón, P. Espinet‑González, Y. Contreras, et al., "Implications of low breakdown voltage of component subcells on external quantum efficiency measurements of multijunction solar cells". *Prog Photovoltaics*, 23, pp 1597-1607, 2015.

[46]   C. Domínguez, I. Antón, and G. Sala, "Multijunction solar cell model for translating I–V characteristics as a function of irradiance, spectrum, and cell temperature". *Prog Photovoltaics*, 18, pp 272-284, 2010.

[47]   C.R. Osterwald, and G. Siefer,"CPV Multijunction Solar Cell Characterization ", in *Handbook of concentrator photovoltaic technology*, John Wiley & Sons, 2016.